\documentclass[12pt]{article}
\usepackage[dvips]{color}
\usepackage{epsfig}
\usepackage{amsmath}
\usepackage{graphicx}

\begin{document}
\title{\bf Energy loss and jet quenching parameter in a thermal non-relativistic non-commutative Yang-Mills plasma}
\author{{J. Sadeghi\thanks{Email: pouriya@ipm.ir}\hspace{1mm} and
B. Pourhassan\thanks{Email: b.pourhassan@umz.ac.ir}}\\
{\small {\em  Sciences Faculty, Department of Physics, Mazandaran University,}}\\
{\small {\em P .O .Box 47416-95447, Babolsar, Iran}}} \maketitle
\begin{abstract}
In this paper, we consider the problem of a moving heavy quark through a hot non-relativistic, non-commutative Yang-Mills plasma. We discuss about the
configuration of the static and dynamic quarks, and also obtain the quasi-normal modes. The main goal of this study is calculating the jet-quenching
parameter for the non-relativistic, non-commutative theory and comparing it with drag forces which recently obtained
from the another independent work [1].
{\bf Pacs Number:} 11.10.Nx ; 11.25.-w
\end{abstract}
\section{Introduction}
Drag force of a heavy moving quark through a hot non-relativistic, non-commutative Yang-Mills plasma has been recently studied by using the AdS/CFT
correspondence [1]. As we know the AdS/CFT correspondence [2-8] is a powerful mathematical tool for simplification of some complicated calculations in the
QCD. However, in reality QCD itself is not directly amenable to this correspondence and permit access to various other interesting strongly-coupled gauge
theories. Already the problem of the drag force has been studied in the ordinary $\mathcal{N}=4$ super Yang-Mills thermal plasma [9-25], and in the
$\mathcal{N}=2$ supergravity theory [26, 27, 28], and in the various other backgrounds [29]. Already we found that moving heavy quark through
$\mathcal{N}=2$ supergravity [30] thermal plasma with non-extremal black hole and finite chemical potential is corresponding to that through the
$\mathcal{N}=4$ super Yang-Mills thermal plasma with near-extremal black hole without chemical potential [26, 27, 28]. The holographic picture of the
moving heavy quark through the thermal QGP (quark gluon plasma) with the momentum $P$, the mass $m$, the constant velocity $v$ and an external force $F$ is
the stretched string in the AdS space. In that case the drag force is given by $\dot{P}=F-\mu P$, where $\mu$ is called the friction coefficient. Another
interesting problem in the strongly coupled plasma is the jet-quenching parameter [22-25, 31-41]. In the ultra-relativistic heavy-ion collisions at LHC or
RHIC, interactions between the high-momentum Parton and the QGP are expected to lead to jet energy loss, which is called the jet quenching. The
jet-quenching parameter provides a measurement of the dispersion of the plasma. The jet-quenching parameter usually calculated by using the perturbation
theory, but by using the AdS/CFT correspondence it is possible to compute the jet-quenching parameter in the non-perturbative quantum field theory. The
perturbative QCD is not very reliable for current experimental temperature, therefore the non-perturbative one have used in recent literatures. It is known
that the non-perturbative definition of the jet-quenching parameter may be obtained in terms of light-like Wilson loop [31]. Fitting the current data it
seems that the value of
the jet-quenching parameter is within the range 5-25 $GeV^{2}/fm$ [39].\\
Also the shear viscosity in the strongly coupled plasma can be calculated by using the AdS/CFT correspondence. In that case universality of the ratio of
the shear viscosity to the entropy density has already been studied in various backgrounds [42-48]. In the Ref. [46] the shear viscosity of the
$\mathcal{N}=2$ supergravity thermal plasma and strong coupling limit of the shear viscosity for the $\mathcal{N}=4$
super-Yang-Mill theory with a chemical potential computed.\\
In the Ref. [1] the non-relativistic non-extremal (D1, D3) bound state solution of type IIB string theory, by using the standard procedure of Null Melvin
Twist [49-52] constructed. A particular low energy limit of this configuration reduced to a non-commutative Yang-Mills theory [53, 54, 55], as well as
coincident non-relativistic (D1, D3) bound state system. In some unified theories, such as great unification theories (GUT), it has been proposed that
space-time coordinates could be non-commutative. So, the non-commutativity is an interesting subject in modern physics. Therefore, we have strong
motivation to study the plasma that simultaneously incorporates non-relativistic and non-commutative features. The origin of the non-commutativity in the
D3-brane is existing the large magnetic field in the background. Also the CFT usually has relativistic nature. However, in the context of condensed matter
systems, it is useful to find holographic descriptions of CFT with non-relativistic nature [56-59]. These systems sometimes can be produced in the
laboratory and indeed there exist such a strongly coupled non-relativistic system such as cold fermions. Therefore, it is interesting to study
non-relativistic, non-commutative QGP. In the Ref. [1], it is found that, in the non-relativistic, non-commutative Yang-Mills theory, explicit expression
of the drag force has not closed form, and the expression of the drag force in various limits has different forms. We deal with this difficulty in
calculating the jet-quenching parameter too. Therefore, we discuss the effect of the non-relativistic and non-commutative nature of the theory for large
and small corresponding parameter. However, up to the constant, we success to find a closed form for the
jet quenching parameter in the relativistic case.\\
In this paper, we completed the discussion of the drag force and quasi-normal modes. An important difference between our work about drag force with the
Ref. [1] is discussion of static quark and zero temperature limit. Moreover, we calculate the jet-quenching parameter in the non-relativistic,
non-commutative
Yang-Mills theory.\\
This paper organized as follows. In section 2 we obtain the equation of motion and momentum densities. We discuss the straightforwardly stretched string at
zero and finite temperatures. In section 3 we study the quasi-normal modes of the string and obtain the lowest modes. In section 4 we compute the
jet-quenching parameter and discuss the effect of non-commutativity parameter and non-relativistic nature of the theory. Finally in section 5 we summarize
our results.
\section{Drag Force}
We begin with the following background metric in the original coordinates [1],
\begin{eqnarray}\label{s1}
ds^{2}&=&\frac{r^{2}}{KR^{2}}\left[(1-\beta^{2}r^{2}f)(dx^{-})^{2}-(1+\beta^{2}r^{2})f(dx^{+})^{2}
+2\beta^{2}r^{2}fdx^{-}dx^{+}\right]\nonumber\\
&+&\frac{hr^{2}}{R^{2}}((dx^{2})^{2}+(dx^{3})^{2})+\frac{R^{2}}{f
r^{2}}dr^{2},
\end{eqnarray}
where we neglected the 5-sphere ($S^{5}$) part of the metric (It has not contribution in our calculations because we calculate in $AdS_{5}$ space only).
Indeed, the metric (1) represents the $AdS_{5}$ space, and $r$ denotes vertical direction to D-branes. In the above solution $K\equiv
1+\beta^{2}\frac{r_{h}^{4}}{r^{2}}$, and $R^{2}=r_{h}^{2}\sinh\varphi$, also $r_{h}$ denotes the horizon radius and $\varphi$ is called the boost
parameter, also $\beta$ is a physical parameter related to the chemical potential of the Yang-Mills theory on the boundary. Moreover,
\begin{eqnarray}\label{s2}
f&=&1-\frac{r_{h}^{4}}{r^{4}},\nonumber\\
h&=&\frac{1}{1+a^{4}r^{4}},
\end{eqnarray}
where,
\begin{equation}\label{s3}
a^{4}=\frac{1}{r_{h}^{4}\sinh^{2}\varphi\cos^{2}\theta}.
\end{equation}
D3-branes are lying along $x^{1}, x^{2}, x^{3}$, and D1-branes are lying along $x^{1}$. The angle $\theta$ in the relation (3) measures the relative numbers of D-branes.
So, for $N$ D3-branes and $M$ D1-branes one can write, $\cos\theta=\frac{N}{\sqrt{N^{2}+M^{2}}}$.\\
In this configuration there is a large magnetic field in the $x^{2}-x^{3}$ directions, so these directions satisfy the non-commutativity relation $[x^{2},
x^{3}]=i\Theta$, where $\Theta$ is called the non-commutativity parameter [53]. It have shown that $a^{4}r_{h}^{4}\sim\Theta^{2}$ [1], so the parameter $a$
measures the non-commutativity. Also $\beta\rightarrow0$ limit recovered the relativistic cases, so the parameter $\beta$ specifies
the non-relativistic feature.\\
Therefore, an external quark moves non-relativistically at
non-commutative plasma. Dual picture of this configuration is the
stretched string from the brane to the horizon. The end point of
string on the brane represents external quark and string may moves
along the non-commutative direction $x^{2}\equiv x$. The open string
is described by the following Nambu-Goto action,
\begin{equation}\label{s4}
S=-T_{0}\int{d\tau d\sigma\sqrt{-g}},
\end{equation}
where $T_{0}$ is the string tension and $(\tau, \sigma)$ are the
string world-sheet coordinates. Also $g$ is the determinant of the
world-sheet metric $g_{ab}$. In the static gauge, $x^{+}\equiv
t=\tau$ and $r=\sigma$, the string world-sheet is described by the
function $x(t, r)$, so the lagrangian density is given by,
\begin{equation}\label{s5}
{\mathcal{L}}=\sqrt{-g}
=\left[\frac{1+\beta^{2}r^{2}}{K}-\frac{h}{f}{\dot{x}}^{2}+\frac{r^{4}}{R^{4}}\frac{1+\beta^{2}r^{2}}{K}h
f{x^{\prime}}^{2}\right]^{\frac{1}{2}},
\end{equation}
where prime and dot denote derivative with respect to $r$ and $t$
respectively. Then, by using the Euler-Lagrange equation one can
find the string equation of motion,
\begin{equation}\label{s6}
\frac{\partial}{\partial
r}\left[\frac{r^{4}}{R^{4}}\frac{1+\beta^{2}r^{2}}{K}h
f\frac{x^{\prime}}{\sqrt{-g}}\right]=\frac{h}{f}\frac{\partial}{\partial
t}\left[\frac{\dot{x}}{\sqrt{-g}}\right].
\end{equation}
In order to obtain the canonical momentum densities associated with
the string we use the following expressions,
\begin{eqnarray}\label{s7}
\pi_{\mu}^{0}&=&-T_{0}G_{\mu\nu}\frac{(\dot{X}\cdot X^{\prime})(X^{\nu})^{\prime}-(X^{\prime})^{2}{\dot{X}}^{\nu}}{\sqrt{-g}}\nonumber\\
\pi_{\mu}^{1}&=&-T_{0}G_{\mu\nu}\frac{(\dot{X}\cdot X^{\prime}){\dot{X}}^{\nu}-(\dot{X})^{2}(X^{\nu})^{\prime}}{\sqrt{-g}},
\end{eqnarray}
where the metric $G_{\mu\nu}$ is given by the relation (1). So, for $\mu, \nu=x, r, t$ one can obtain,
\begin{eqnarray}\label{s8}
\left(\begin{array}{ccc}
\pi_{x}^{0} & \pi_{x}^{1}\\
\pi_{r}^{0} & \pi_{r}^{1}\\
\pi_{t}^{0} & \pi_{t}^{1}\\
\end{array}\right)=-\frac{T_{0}}{\sqrt{-g}} \left(\begin{array}{ccc}
-\frac{h}{f}\dot{x} & \frac{r^{4}}{R^{4}}\frac{1+\beta^{2}r^{2}}{K}h
fx^{\prime}\\
\frac{h}{f}\dot{x}x^{\prime} & \frac{1+\beta^{2}r^{2}}{K}-\frac{h}{f}{\dot{x}}^{2}\\
\frac{1+\beta^{2}r^{2}}{K}(1+\frac{r^{4}}{R^{4}}h f{x^{\prime}}^{2})
& \frac{r^{4}}{R^{4}}\frac{1+\beta^{2}r^{2}}{K}h f
\dot{x}x^{\prime}\\
\end{array}\right).
\end{eqnarray}
Relation (8) is the general expression of canonical momentum of the string stretched from $r=r_{m}$ on the brane to $r=r_{h}$. In that case the total
energy and momentum of the string are calculated by the following relations respectively,
\begin{eqnarray}\label{s9}
E&=&-\int_{r_{h}}^{r_{m}} dr\pi_{t}^{0},\nonumber\\
P&=&\int_{r_{h}}^{r_{m}} dr\pi_{x}^{0}.
\end{eqnarray}
The component $\pi_{x}^{1}$ interpreted as the drag force on the
quark due to thermal plasma. The simplest solution of the equation
of motion (6) is $x=x_{0}$, where $x_{0}$ is a constant. In this
case the string stretched straightforwardly from $r=r_{m}$ to the
$r=r_{h}$. This configuration is dual picture of the static quark in
the thermal non-commutative plasma. In this case
$-g=\frac{1+\beta^{2}r^{2}}{K}$, and drag force vanishes which is
expected for the static quark. Only non-zero components of the
momentum density obtained as,
\begin{equation}\label{s10}
\pi_{r}^{1}=\pi_{t}^{0}=-T_{0}\sqrt{\frac{1+\beta^{2}r^{2}}{K}}.
\end{equation}
By using the first relation of the (9) one can obtain the total
energy of the string. For the case of $\beta\ll1$ we get,
\begin{equation}\label{s11}
E=T_{0}\left[r_{m}-r_{h}+ \frac{\beta^{2}}{2}(\frac{r_{h}^{4}}{r_{m}}+\frac{r_{m}^{3}}{3}-\frac{4r_{h}^{2}}{3})\right]+\mathcal{O}(\beta^{4}).
\end{equation}
It is clear that at $\beta\rightarrow0$ limit we recover the result of the Ref. [9], where the total energy of the string obtained as $E\sim r_{m}-r_{h}$.
Therefore, $\beta\rightarrow0$ limit of these calculations is corresponding to the moving heavy quark through
$\mathcal{N}=4$ super Yang-Mills plasma.\\
The temperature of the non-relativistic, non-commutative Yang-Mills
theory is given by [1],
\begin{equation}\label{s12}
T=\frac{r_{h}}{\pi R^{2}}=\frac{r_{h}}{\sqrt{2\hat{\lambda}}\pi
\alpha^{\prime}},
\end{equation}
where $\alpha^{\prime}$ is the slop parameter $(\alpha^{\prime}=\frac{1}{2\pi T_{0}})$, and $\hat{\lambda}$ is the 't Hooft coupling of the
non-relativistic, non-commutative theory, which is related to the 't Hooft coupling of the ordinary Yang-Mills theory by the relation
$\lambda=\frac{\alpha^{\prime}}{\Theta}\hat{\lambda}$. Therefore, zero-temperature limit obtained by taking $r_{h}\rightarrow0$ limit. At the
zero-temperature limit one can interpret $E$ as the physical mass of the quark, hence in this limit one can obtain,
\begin{equation}\label{s13}
(E)_{T=0}=m=T_{0}r_{m}\left[1+\frac{\beta^{2}r_{m}^{2}}{6}\right].
\end{equation}
As we expected at the $r_{m}\rightarrow\infty$ limit (brane position moves to the boundary) the quark mass will be infinite. Also, increasing the
temperature decreases the quark mass, so one can write,
\begin{equation}\label{s14}
(E)_{T\neq0}=M_{rest}(T)=m-\Delta m(T),
\end{equation}
where thermal rest mass shift is defined as,
\begin{equation}\label{s15}
\Delta m(T)=E=T_{0}r_{h}\left[1+\frac{\beta^{2}}{2}(4\frac{r_{h}}{3}-\frac{r_{h}^{3}}{r_{m}})\right].
\end{equation}
We summarized these results in the table 1.

\begin{center}
  \begin{tabular}{|@{} l @{} ||@{} c @{} | @{}r @{}|}
    \hline
    Quantity & NR-NC YM & Type IIB string \\ \hline\hline
    Slope parameter & $\frac{R^{2}}{\sqrt{2\hat{\lambda}}}$ & $\alpha^{\prime}$ \\ \hline
    't Hooft coupling & $\hat{\lambda}$ & $\frac{R^{4}}{2\alpha^{\prime 2}}$ \\ \hline
    Temperature & $T$ & $\frac{r_{h}}{\pi R^{2}}$ \\ \hline
    Horizon radius & $\pi R^{2} T$ & $r_{h}$  \\ \hline
    Thermal rest mass shift & $\Delta m(T)$ & $T_{0}r_{h}\left[1+\frac{\beta^{2}}{2}r_{h}(\frac{4}{3}-\frac{r_{h}^{2}}{r_{m}})\right]$  \\ \hline
    Physical mass & $m$ & $T_{0}r_{m}\left[1+\frac{\beta^{2}r_{m}^{2}}{6}\right]$  \\ \hline
    Static thermal mass & $M_{rest}(T)$ & $T_{0}\left[r_{m}-r_{h}+\frac{\beta^{2}}{2}(\frac{r_{h}^{4}}{r_{m}}+\frac{r_{m}^{3}}{3}-\frac{4r_{h}^{2}}{3})\right]$ \\
    \hline
  \end{tabular}
\end{center}
Table 1. AdS/CFT  translation table. Expressions of $\Delta m(T)$,
$M_{rest}(T)$ and $m$ obtained for the infinitesimal $\beta$ and
$\mathcal{O}(\beta^{4})$ neglected. These results are agree with
[9] at $\beta\rightarrow0$ limit.\\\\\\
In the other hand for the case of $\beta\gg1$ one can obtain,
\begin{equation}\label{s16}
E\approx T_{0}\frac{\beta^{2}}{2}r_{h}\left[r_{m}-r_{h}-\beta r_{h}\ln{\frac{r_{m}+\beta r_{h}}{r_{h}+\beta r_{h}}}\right].
\end{equation}
In this cases the non-commutative parameter $a$ is not important because the static quark has no motion along the non-commutative
directions.\\
Another solution of the equation of motion (6) may be written as $x=vt+x_{0}$, it is corresponding to the moving straightforward
string which is not physical solution [9, 26].\\
The other solution which satisfies the equation of motion may be
written as $x(t, r)=vt+\xi(r)$. Such a solution considered recently
[1] and the drag force obtained in the standard form as the
following,
\begin{equation}\label{s17}
\pi_{x}^{1}=-T_{0}Cvr_{c}^{2},
\end{equation}
where constant $C$ is given by,
\begin{equation}\label{s18}
C=\frac{1}{r_{h}^{2}(1+a^{4}r_{c}^{4})\sinh\varphi},
\end{equation}
and the critical radius $r_{c}$ is the root of the following
equation,
\begin{equation}\label{s19}
(r^{4}-r_{h}^{4})(1+a^{4}r^{4})(1+\beta^{2}r^{2})-v^{2}(r^{4}+\beta^{2}r_{h}^{4}r^{2})=0.
\end{equation}
The expression (17) for the drag force is in agreement with the
previous works such as [9, 26, 27, 28], only differences are
definition of the constant $C$ and the critical radius $r_{c}$. In
the special case of $\beta=a=0$ one can find,
\begin{equation}\label{s20}
\pi_{x}^{1}=-T_{0}\frac{v}{\sqrt{1-v^{2}}\sinh\varphi}.
\end{equation}
Particularly, if we set $\sinh\varphi=-\sqrt{1-v^{2}}$, this result is coincides with [26] at $\eta\rightarrow0$ limit and $\Lambda^{2}=1$, where $\eta$ is
called the non-extremality parameter and $\Lambda$ denotes the cosmological constant. As we know the non-extremality parameter is related to the black hole
charge $q$ [28], and the black hole charge $q$ is related to the chemical potential $(\eta\sim q\sim \frac{1}{\beta})$. So, it is reasonable that results
of heavy quark in non-relativistic, non-commutative Yang-Mills plasma at $\beta\rightarrow0$ and $a\rightarrow0$ limits are agree with the results of the
moving heavy quark through $\mathcal{N}=2$ supergravity thermal plasma at extremal limit without $B$-field [26, 27, 28]. Both theories at mentioned limits
are corresponding to moving heavy quark through $\mathcal{N}=4$ super Yang-Mills thermal plasma without the chemical potential.
\section{Quasi-Normal Modes}
In this section we would like to consider behavior of the curved string at the late time and in the low velocity limit. In that case the string has small
fluctuations around the straight string. It means that ${\dot{x}}^{2}$ and ${x^{\prime}}^{2}$ are infinitesimal, so one can neglect them in the expression
(5). Therefore, the equation of motion reduces to the following equation,
\begin{equation}\label{s21}
\frac{\partial}{\partial
r}\left[\frac{r^{4}}{R^{4}}\sqrt{\frac{1+\beta^{2}r^{2}}{K}} f
x^{\prime}\right]=\frac{1}{f}\sqrt{\frac{K}{1+\beta^{2}r^{2}}}\ddot{x},
\end{equation}
where $a=0$ (there are no movements along the non-commutative
direction). Then, one may choose the time-dependent solution of the
form,
\begin{equation}\label{s22}
x(r, t)=\xi(r) e^{-\mu t}.
\end{equation}
In that case the equation of motion (21) reduces to the following
differential equation,
\begin{equation}\label{s23}
O\xi(r)=\mu^{2}\xi(r),
\end{equation}
where we define,
\begin{equation}\label{s24}
O\equiv f\sqrt{\frac{1+\beta^{2}r^{2}}{K}}\frac{\partial}{\partial
r}\frac{r^{4}}{R^{4}} f
\sqrt{\frac{1+\beta^{2}r^{2}}{K}}\frac{\partial}{\partial r}.
\end{equation}
The anstaz (22) satisfies Neumann boundary condition at $r=r_{m}$
$(\xi^{\prime}(r_{m})=0)$. In order to obtain the friction
coefficient $\mu$, it is convenient to expand $\xi(r)$ as power
series of $\mu$,
\begin{equation}\label{s25}
\xi(r)=\xi_{0}(r)+\mu\xi_{1}(r)+\mu^{2}\xi_{2}(r)+\ldots.
\end{equation}
Substituting expansion (25) in the equation (23) tells that,
\begin{eqnarray}\label{s26}
O\xi_{0}&=&0,\nonumber\\
O\xi_{1}&=&0,\nonumber\\
O\xi_{2}&=&\xi_{0}.
\end{eqnarray}
Neumann boundary condition causes to choose $\xi_{0}=A$, where $A$ is a constant, therefore,
\begin{eqnarray}\label{s27}
\xi^{\prime}(r_{m})=\mu\xi_{1}^{\prime}(r_{m})+\mu^{2}\xi_{2}^{\prime}(r_{m})=0.
\end{eqnarray}
At the $\beta\rightarrow0$ limit, by using the equation (27), one
can obtain,
\begin{eqnarray}\label{s28}
\mu=\left[r_{m}+\frac{r_{h}}{4}\ln{\frac{r_{m}-r_{h}}{r_{m}+r_{h}}}-\frac{r_{h}}{2}\tan^{-1}\frac{r_{m}}{r_{h}}\right]^{-1}.
\end{eqnarray}
On the other hand, for $\beta\neq0$, by using the near horizon behavior ($r\rightarrow r_{h}$ approximation which yields $f\approx1$ and
$\frac{K}{1+\beta^{2}r^{2}}\approx1$), one can obtain,
\begin{eqnarray}\label{s29}
\xi_{1}^{\prime}(r)&=&-A\frac{R^{4}}{r^{4}}\nonumber\\
\xi_{2}^{\prime}(r)&=&A\frac{R^{4}}{r^{3}}.
\end{eqnarray}
Therefore, applying the Neumann boundary condition yields to the following expression for the friction coefficient,
\begin{equation}\label{s30}
\mu=\frac{1}{r_{m}},
\end{equation}
which is agree with the first term of the expression (28). This is
the smallest eigenvalue of operator $O$ which obtained for the given
boundary condition.\\
As mentioned above, for the infinitesimal $\dot{x}$ and $x^{\prime}$ one can write $\sqrt{-g}\approx\sqrt{\frac{1+\beta^{2}r^{2}}{K}}$, then the momentum
density is given by,
\begin{equation}\label{s31}
\pi_{x}^{0}=-\frac{T_{0}}{\mu}\left[\frac{r^{4}}{R^{4}} f
\sqrt{\frac{1+\beta^{2}r^{2}}{K}}x^{\prime}\right]^{\prime},
\end{equation}
where we used the equation of motion (21) and the solution (22).
Then, we use the second relation (9) to obtain the total momentum of
the string,
\begin{equation}\label{s32}
P=\frac{T_{0}}{\mu}\left[\frac{r_{min}^{4}}{R^{4}}
(1-\frac{r_{h}^{4}}{r_{min}^{4}})
\sqrt{\frac{1+\beta^{2}r_{min}^{2}}{1+\beta^{2}\frac{r_{h}^{4}}{r_{min}^{2}}}}x^{\prime}(r_{min})\right],
\end{equation}
where we used the Neumann boundary condition, and $r_{min}>r_{h}$ as
an IR cutoff at a lower limit of the integral.\\
In order to obtain the total energy we expand $\sqrt{-g}$ to the
second order of $\dot{x}$ and $x^{\prime}$, and obtain,
\begin{equation}\label{s33}
\pi_{t}^{0}=-T_{0}\left[\sqrt{\frac{1+\beta^{2}r^{2}}{K}}+\frac{1}{2}\left(\frac{r^{4}}{R^{4}}
f
\sqrt{\frac{1+\beta^{2}r^{2}}{K}}xx^{\prime}\right)^{\prime}\right],
\end{equation}
where we used the equation of motion (21). Then, we use the first
relation (9) to obtain the total energy of the string,
\begin{eqnarray}\label{s34}
E&=&T_{0}\left[r_{m}-r_{min}+\frac{\beta^{2}}{2}(\frac{r_{h}^{4}}{r_{m}}+\frac{r_{m}^{3}}{3}-\frac{4r_{h}^{2}}{3})
\right]\nonumber\\
&+&\frac{T_{0}}{2}\frac{r_{min}^{4}}{R^{4}}
(1-\frac{r_{h}^{4}}{r_{min}^{4}})
\sqrt{\frac{1+\beta^{2}r_{min}^{2}}{1+\beta^{2}\frac{r_{h}^{4}}{r_{min}^{2}}}}x(r_{min})x^{\prime}(r_{min}),
\end{eqnarray}
where we assume that the parameter $\beta$ is infinitesimal and used
the Neumann boundary condition. Combining the relations (32) and
(34) and using $\dot{x}=-\mu mx$ yields to the simple relationship
$E=M_{rest}+\frac{P^{2}}{2m}$, where $m$ is the kinetic mass of the
quark.
\section{Jet-Quenching Parameter}
In order to calculate the jet-quenching parameter we should consider an open string whose endpoints lie on the brane. This is corresponding to
quark-antiquark configuration. In the light cone coordinates the string profile is given by the function $r(t, y)$, where we used static gauge,
$\tilde{x}^{-}\equiv\tau=t \hspace{3mm}(L^{-}\leq \tilde{x}^{-}\leq0)$ and $x^{2}\equiv \sigma=y\hspace{3mm} (-\frac{L}{2}\leq y\leq \frac{L}{2})$, and all
other coordinates are constant. Because of the condition $L^{-}\gg L$, the world-sheet is invariant along the $\tilde{x}^{-}$ direction, and one can
consider the function $r(y)$ as the string profile, so $r(\pm\frac{L}{2})=\infty$. Also we use the light cone coordinates
$\tilde{x}^{-}=\frac{1}{\sqrt{2}}(x^{+}-x^{-})$ and $\tilde{x}^{+}=\frac{1}{\sqrt{2}}(x^{+}+x^{-})$ in the metric (1) [1, 56, 57, 58, 59]. In that case the
Nambu-Goto action (4) takes the following form,
\begin{equation}\label{s35}
S=2T_{0}L^{-}\int_{0}^{\frac{L}{2}}dy\sqrt{\frac{(\frac{r_{h}^{4}}{2r^{4}}-2r^{2}\beta^{2}f)}{K}(\frac{r^{4}}{R^{4}}h+\frac{r^{\prime
2}}{f})},
\end{equation}
where prime denotes derivative with respect to $y$. Equation of motion ($\mathcal{H}=\mathcal{L}-\frac{\partial \mathcal{L}}{\partial
r^{\prime}}r^{\prime}=\varepsilon$) yields to the following expression,
\begin{equation}\label{s41}
r^{\prime
2}=\frac{fh}{\varepsilon^{2}}\frac{r^{4}}{R^{4}}\left[\frac{(\frac{r_{h}^{4}}{2r^{4}}-2r^{2}\beta^{2}f)}{K}\frac{r^{4}}{R^{4}}
h-\varepsilon^{2}\right],
\end{equation}
where we interpreted the constant $\varepsilon$ as the string energy. By using the relation (36) in (35) and $dy=dr/r^{\prime}$ one can rewrite the action
(35) as the following,
\begin{equation}\label{s37}
S=2T_{0}L^{-}\int_{r_{h}}^{\infty}dr\sqrt{\frac{(\frac{r_{h}^{4}}{2r^{4}}-2r^{2}\beta^{2}f)}{fK}}
\left(1-\frac{\varepsilon^{2}}{(\frac{r_{h}^{4}}{2r^{4}}-2r^{2}\beta^{2}f)}\frac{R^{4}}{r^{4}}\frac{K}{h}\right)^{-\frac{1}{2}}.
\end{equation}
For the low energy limit $(\varepsilon\ll1)$, which is corresponding
to $L\ll L^{-}$, one can obtain,
\begin{equation}\label{s38}
S=S_{0}+
T_{0}L^{-}\varepsilon^{2}\int_{r_{h}}^{\infty}dr\frac{R^{4}}{r^{4}}\sqrt{\frac{K}{fh^{2}(\frac{r_{h}^{4}}{2r^{4}}-2r^{2}\beta^{2}f)}},
\end{equation}
where
\begin{equation}\label{s39}
S_{0}=
2T_{0}L^{-}\int_{r_{h}}^{\infty}dr\sqrt{\frac{fh^{2}(\frac{r_{h}^{4}}{2r^{4}}-2r^{2}\beta^{2}f)}{fK}},
\end{equation}
interpreted as self energy of the isolated quark and antiquark. Also
one can integrate the equation (36) and obtain the following
relation,
\begin{equation}\label{s40}
L=\varepsilon\int_{r_{h}}^{\infty}dr\frac{R^{4}}{r^{4}}\sqrt{\frac{K}{fh^{2}(\frac{r_{h}^{4}}{2r^{4}}-2r^{2}\beta^{2}f)}}.
\end{equation}
Therefore, one can find,
\begin{equation}\label{s41}
S-S_{0}= T_{0}L^{-}L^{2}
\left[\int_{r_{h}}^{\infty}dr\frac{R^{4}}{r^{4}}\sqrt{\frac{K}{fh^{2}(\frac{r_{h}^{4}}{2r^{4}}-2r^{2}\beta^{2}f)}}\right]^{-1}.
\end{equation}
Finally by using the following relation [25],
\begin{equation}\label{s42}
\hat{q}\equiv2\sqrt{2}\frac{S-S_{0}}{L^{-}L^{2}},
\end{equation}
we find the expression of the jet-quenching parameter,
\begin{equation}\label{s43}
\hat{q}=2\sqrt{2}T_{0}
\left[\int_{r_{h}}^{\infty}dr\frac{R^{4}}{r^{4}}\sqrt{\frac{K}{fh^{2}(\frac{r_{h}^{4}}{2r^{4}}-2r^{2}\beta^{2}f)}}\right]^{-1}
\end{equation}
Before anything else we check the validity of the above expression
at $\beta\rightarrow0$ and $a\rightarrow0$ limits. In these limits
one find $K=1$ and $h=1$, so one can obtain,
\begin{equation}\label{s44}
\hat{q}_{SYM}=\frac{\pi^{2}}{b}\sqrt{\lambda}T^{3},
\end{equation}
where we used $T_{0}=\frac{1}{2\pi\alpha^{\prime}}$,
$T=\frac{r_{h}}{\pi R^{2}}$ and
$R^{2}=\alpha^{\prime}\sqrt{\lambda}$. Also
$b=\sqrt{\pi}\frac{\Gamma(\frac{5}{4})}{\Gamma(\frac{3}{4})}\approx1.311$.
The equation (44) is the well known relation of the jet-quenching
parameter in the hot $\mathcal{N}=4$ supersymmetric QCD [31].\\
From the Ref. [1] it is found that, for the case of $\beta=0$ and
$a=0$ the drag force $\dot{P}$ proportional to $\sqrt{\lambda}\pi
T^{2}v$. Therefore in this case one can obtain,
\begin{equation}\label{s45}
(\frac{\hat{q}}{\dot{P}})_{SYM}\sim T.
\end{equation}
It means that the ratio of the jet-quenching parameter to the drag
force is linear for the temperature in the $\mathcal{N}=4$ super
Yang-Mills plasma without the chemical potential.\\
One can consider another case with $\beta\rightarrow0$ and $ar_{h}\gg1$ which yields to the following expression,
\begin{equation}\label{s46}
\hat{q}_{NC SYM}=\frac{12\pi^{2}}{5}\sqrt{2\hat{\lambda}}T^{3},
\end{equation}
where we assume $a^{2}r_{h}^{2}=\Theta$. This assumption is consistent with the relation (3) with $\Theta=(\sinh\varphi\cos\theta)^{-1}$. The large
non-commutativity parameter $a$ means that $r_{h}$ will be small parameter, so we neglect $\mathcal{O}(r_{h}^{4})$ terms in the above relation. Therefore,
the value of the temperature in the relation (46) is lower than the value of the temperature in the relation (44), hence it seems that the jet-quenching
parameter for the case of $\beta\rightarrow0$, $a\gg1$ is smaller than the jet-quenching parameter for the case of $\beta\rightarrow0$ and
$ar_{h}\rightarrow0$. But, we should note that $\hat{\lambda}>\lambda$ for large non-commutativity parameter. So, in order to compare $\hat{q}_{NC SYM}$
with $\hat{q}_{SYM}$ we exam value of them at $T=300 MeV$. In that case one can obtain $\hat{q}_{SYM}\approx4.5$ $GeV^{2}/fm$, and $\hat{q}_{NC
SYM}\approx19$ $GeV^{2}/fm$, which is nicely in the experimental range [39]. Therefore, presence of the non-commutativity increases the value of the
jet-quenching parameter. In order to obtain numerical value of $\hat{q}_{NC SYM}$, we assumed $a$ of order $10^{5}$, which is consistent with value of
$\Theta$ in the Ref. [60], to find a significant correction due to non-commutativity in collider experiments. Such value is also agree with spin statistics
violations in non-commutative QED from Gran Sasso and Super-Kamiokande [61]. Also, we can impose other limits on non-commutativity from experiments, for
examples quantum mechanics, Lamb shift in non-commutative QED and non-commutative extensions of standard model give $a\sim10$ [62], and non-commutative
symplectic structure in classical mechanics and perihelion of Mercury give $a\sim10^{13}$ [63].

In the both cases we find standard form of the jet-quenching parameter proportional to $T^{3}$ times square root of the 't Hooft coupling. Although the
drag force for the large non-commutativity parameter is proportional to $(\sqrt{\hat{\lambda}}T^{2})^{-1}$, but the jet-quenching parameter saves its
shape. However authors in the Ref. [1], for the large non-commutativity, concluded that the drag force on quark is very small and this result is in
agrement with our result.\\
The next case which we consider in this paper is the case of
$\beta\gg1$ at $a\rightarrow0$ limit. In this case one can obtain,
\begin{equation}\label{s47}
\hat{q}_{NR SYM}=\frac{\bar{\mu}^{2}}{T\sqrt{2\hat{g}_{YM}^{2}N}}I^{-1},
\end{equation}
where $\bar{\mu}=(\beta\alpha^{\prime})^{-1}$ defined as the chemical potential of the non-relativistic, non-commutative Yang-Mills theory, and we defined,
\begin{equation}\label{s48}
I=\int_{r_{h}}^{\infty}\frac{dr}{r\sqrt{(r^{4}-r_{h}^{4})\beta^{2}(r_{h}^{4}-4r^{2}\beta^{2}(r^{4}-r_{h}^{4}))}}.
\end{equation}
In the case of high temperature $(r_{h}\rightarrow\infty)$ one can
obtain $I\propto \beta^{-2}r_{h}^{-4}$. Therefore, we found that
$\hat{q}_{NR SYM}\propto T$, comparing the jet-quenching parameter
with the drag force yields to the following relation,
\begin{equation}\label{s49}
(\frac{\hat{q}}{F})_{NR SYM}\sim T.
\end{equation}
In the Ref. [1] it is found that the drag force for the case of
$\beta\gg1$ and $ar_{h}\ll1$ dose not depend on the temperature. It
means that the ratio of the jet-quenching parameter to the drag
force at high temperature limit is proportional to the temperature
which already obtained for the case of ordinary theory [see relation
(45)].\\
Finally, we consider the case of $\beta\ll1$ and $ar_{h}\ll1$. So, we obtain,
\begin{equation}\label{s50}
\hat{q}_{NC NR
SYM}\propto\frac{\pi^{2}a^{2}\sqrt{2\hat{\lambda}}T^{3}}{a^{2}+\pi^{2}\hat{\lambda}\alpha^{\prime2}\beta^{2}T^{2}}.
\end{equation}
It is clear that $a\rightarrow0$ limit of the equation (50) yields
to $\hat{q}_{NR SYM}\propto T$ and $\beta\rightarrow0$ limit of the
equation (50) yields to $\hat{q}_{NC SYM}\propto T^{3}$, which agree
with the results of the relations (45) and (47).
\section{Conclusion}
In this paper we considered non-relativistic, non-commutative Yang-Mills plasma and studied the problem of the moving heavy quark through the thermal
plasma. As we mentioned in the introduction, the non-relativistic nature of CFT is important for the condensed matter theory. Also existing the large
magnetic field in the background yields to non-commutativity in the background which is important for some unified theories. Therefore, study of
non-relativistic non-commutative QGP is interesting. We obtained the full components of the momentum density and discussed about the static quark
configuration. Then, we discussed about the quasi-normal modes. Finally we computed the jet-quenching parameter for the non-relativistic, non-commutative
theory. We concluded for the large chemical potential (for both infinitesimal and large non-commutativity parameter) the jet-quenching parameter obtained
in its standard form, but for the case of infinitesimal non-commutativity parameter and large $\beta$ the jet quenching parameter obtained proportional to
the temperature. However, in this case, the ratio of the jet-quenching parameter to the drag force is similar to the ordinary theory. We found that the
presence of non-commutativity is necessary to obtain the jet-quenching parameter in the experimental range. As we know the experimental data tell that the
value of the jet-quenching should be in the range $(15\pm10)GeV^{2}/fm$ [39]. We found that $\hat{q}_{NC SYM}\approx19GeV^{2}/fm$ at $T=300$ $MeV$ which is
in the experimental range. We should note that our results differ with the usual results obtained by means of calculation in quantum field theory involving
the usual mechanisms of multiple scattering and radiative energy loss [64]. For example the value of the jet-quenching parameter obtained about 2.3
$GeV^{2}/fm$ by using multiple scattering mechanism for $T=400$ $MeV$ which is clearly lower than our result and experimental data. The reason is that the
AdS/CFT correspondence gives more exact solutions than other methods.


\begin{thebibliography}{11}
\bibitem{P1}
Kamal L. Panigrahi, Shibaji Roy, "Drag force in a hot non-relativistic, non-commutative Yang-Mills plasma", JHEP04(2010)003, [arXiv:1001.2904 [hep-th]].
\bibitem{P2}
J. M. Maldacena, "The large N limit of superconformal field theories
and supergravity", Adv. Theor. Math. Phys. \textbf{2} (1998) 231.
\bibitem{P3}
E. Witten, "Anti-de Sitter space and holography", Adv. Theor. Math.
Phys. \textbf{2} (1998) 253.
\bibitem{P4}
S. S. Gubser, I. R. Klebanov, and A. M. Polyakov, "Gauge theory
correlators from noncritical string theory", Phys. Lett.
\textbf{B428} (1998) 105.
\bibitem{P5}
J. H. Schwart, "Introduction to M Theory and $AdS$/CFT Duality", Lecture Notes in Physics, Volume 525(1999), [arXiv:hep-th/9812037].
\bibitem{P6}
Juan Maldacena, "TASI 2003 lectures on AdS/CFT", [arXiv:hep-th/0309246]. M. R. Douglas and S. Randjbar-Daemi, "Two Lectures on $AdS$/CFT correspondence"
[arXiv:hep-th/9902022].
\bibitem{P7}
J. L. Petersen, "Introduction to the Maldacena Conjecture on $AdS$/CFT", Int. J. Mod. Phys. \textbf{A14} (1999) 3597.
\bibitem{P8}
Horatiu Nastase, "Introduction to AdS-CFT", arXiv:0712.0689
[hep-th]. Laurent Freidel, "Reconstructing AdS/CFT",
[arXiv:0804.0632 [hep-th]].
\bibitem{P9}
C. P. Herzog, A. Karch, P. Kovtun, C. Kozcaz, and L. G. Yaffe,
"Energy loss of a heavy quark moving through ${\mathcal{N}} =4$
supersymmetric Yang-Mills plasma" JHEP \textbf{0607} (2006) 013.
\bibitem{P10}
C. P. Herzog, "Energy loss of heavy quarks from asymptotically AdS
geometries", JHEP \textbf{0609} (2006) 032.
\bibitem{P11}
S. S. Gubser, "Drag force in AdS/CFT", Phys. Rev. \textbf{D74}
(2006) 126005.
\bibitem{P12}
E. Caceres and A. Guijosa, "Drag force in charged ${\mathcal{N}} =4$
SYM plasma". JHEP \textbf{0611} (2006) 077.
\bibitem{P13}
T. Matsuo, D. Tomino and W. Y. Wen, "Drag force in  SYM plasma with
$B$ field from $AdS$/CFT", JHEP \textbf{0610} (2006) 055.
\bibitem{P14}
J. F. Vazquez-Poritz, "Drag force at finite 't Hooft coupling from
$AdS$/CFT" [arxiv:0803.2890 [hep-th]].
\bibitem{P15}
M. Chernicoff, J. A. Garcia and A. Guijosa, "The energy of a moving quark-antiquark pair in an ${\mathcal{N}} =4$ SYM plasma", JHEP \textbf{0609} (2006)
068. M. Chernicoff, D. Fernandez, D. Mateos and D. Trancanelli, "Jet quenching in a strongly coupled anisotropic plasma", [arXiv:1203.0561 [hep-th]].
\bibitem{P16}
J. J. Friess, S. S. Gubser and G. Michalogiorgakis, "Dissipation
from a heavy quark moving through N = 4 super-Yang-Mills plasma",
JHEP \textbf{0609} (2006) 072.
\bibitem{P17}
K. Peeters and M. Zamaklar, "The string/gauge theory correspondence
in QCD", Eur. Phys. J. Special Topics \textbf{152} (2007) 113.
\bibitem{P18}
J. Erdmenger, N. Evans, I. Kirsch, and E. J. Threlfall, " Mesons in
gauge/gravity duals", Eur. Phys. J. A \textbf{35} (2008) 81.
\bibitem{P19}
J. Sadeghi, S. Heshmatian, "Screening Length of Rotating Heavy Meson
from AdS/CFT", Int J Theor Phys (2010) 49: 1811, [arXiv:0812.4816
[hep-th]].
\bibitem{P20}
K. B. Fadafan, "$R^{2}$ curvature-squared corrections on drag force", JHEP 0812 (2008) 051, [arXiv:0803.2777 [hep-th]].
\bibitem{P21}
M. Ali-Akbari, K. Bitaghsir Fadafan, "Rotating mesons in the presence of higher derivative corrections from gauge-string duality", Nucl. Phys. B835 (2010)
221,  [arXiv:0908.3921 [hep-th]].
\bibitem{P22}
K. B. Fadafan, "Medium effect and finite 't Hooft coupling correction on drag force and Jet Quenching Parameter", [arXiv:0809.1336 [hep-th]]. K. Bitaghsir
Fadafan, B. Pourhassan, and J. Sadeghi, "Calculating the jet-quenching parameter in STU background", Eur. Phys. J. C 71 (2011) 1785, [arXiv:1005.1368
[hep-th]].
\bibitem{P23}
E. Nakano, S. Teraguchi and W. Y. Wen, "Drag Force, Jet Quenching,
and AdS/QCD", Phys. Rev. \textbf{D75} (2007) 085016.
\bibitem{P24}
E. Caceres and A. Guijosa, "On drag forces and jet quenching in
strongly coupled plasmas", JHEP \textbf{0612} (2006) 068.
\bibitem{P25}
G. Bertoldi, F. Bigazzi, A. L. Cotrone, Jose D. Edelstein, "Holography and Unquenched Quark-Gluon Plasmas", Phys. Rev. D76 (2007) 065007,
[arXiv:hep-th/0702225].
\bibitem{P26}
J. Sadeghi and B. Pourhassan, "Drag Force of Moving Quark at The $\mathcal{N}=2$ Super-gravity", JHEP 0812 (2008) 026, [arXiv:0809.2668 [hep-th]].
\bibitem{P27}
J. Sadeghi, M. R. Setare, B. Pourhassan and S. Hashmatian, "Drag
Force of Moving Quark in STU Background", Eur. Phys. J. C 61 (2009)
527, [arXiv:0901.0217 [hep- th]].
\bibitem{P28}
J. Sadeghi, M. R. Setare and B. Pourhassan, "Drag force with different charges in STU background and AdS/CFT", J. Phys. G: Nucl. Part. Phys. 36 (2009)
115005, [arXiv:0905.1466 [hep-th]].
\bibitem{P29}
K. B. Fadafan, "Drag force in asymptotically Lifshitz spacetimes", arXiv:0912.4873 [hep-th]. Carlos Hoyos-Badajoz, "Drag and jet quenching of heavy quarks
in a strongly coupled $N=2^{*}$ plasma", JHEP 0909(2009)068, [arXiv:0907.5036 [hep-th]].
\bibitem{P30}
 K. Behrndt, M. Cvetic, W. A. Sabra, "Non-Extreme Black Holes of Five Dimensional N=2 AdS
 Supergravity", Nucl. Phys. B553 (1999) 317, [arXiv:hep-th/9810227].  M. Cvetic, H. Lu, C.N.
 Pope, "Charged Rotating Black Holes in Five Dimensional $U(1)^3$ Gauged N=2
 Supergravity", Phys. Rev. D70 (2004) 081502, [arXiv:hep-th/0407058].  Mirjam Cvetic, Steven S.
 Gubser, "Phases of R-charged Black Holes, Spinning Branes and Strongly Coupled Gauge
 Theories", JHEP 9904 (1999) 024, [arXiv:hep-th/9902195].
\bibitem{P31}
H. Liu, K. Rajagopal and U. A. Wiedemann, "Calculating the jet
quenching parameter from AdS/CFT", Phys. Rev. Lett. \textbf{97}
(2006) 182301.
\bibitem{P32}
Philip C. Argyres, Mohammad Edalati, Justin F. Vazquez-Poritz,
"Spacelike strings and jet quenching from a Wilson loop"
JHEP0704,(2007) 049.
\bibitem{P33}
Philip C. Argyres, Mohammad Edalati, Justin F. Vazquez-Poritz,
"Lightlike Wilson loops from AdS/CFT" JHEP0803 (2008) 071.
\bibitem{P34}
A. Buchel, " On jet quenching parameteres in strongly coupled
non-conformal gauge theories", Phys. Rev. \textbf{D74} (2006)
046006. J. Sadeghi and B. Pourhassan, "Jet-Quenching of the Rotating
Heavy Meson in a N = 4 SYM Plasma in Presence of a Constant Electric
Field", Int J Theor Phys (2011) 50:2305, [arXiv:1001.0706 [hep-th]].
\bibitem{P35}
J. F. Vazquez-Poritz, "Enhancing the jet quenching parameter from marginal deformations", [arXiv:hep-th/0605296].
\bibitem{P36}
F. L. Lin and T. Matsuo, "Jet quenching parameter in medium with
chemical potential from AdS/CFT", Phys. Lett. \textbf{B641} (2006)
45.
\bibitem{P37}
S. D. Avramis and K. Sfetsos, "Supergravity and the jet quenching
parameter in the presence of $R$-charge densities", JHEP
\textbf{0701} (2007) 065.
\bibitem{P38}
N. Armesto, J. D. Edelstein and J. Mas, "Jet quenching at finite 't
Hooft coupling and chemical potential from AdS/CFT", JHEP
\textbf{0609} (2006) 039.
\bibitem{P39}
J. D. Edelstein and C. A. Salgado, "Jet quenching in heavy Ion collisions from AdS/CFT", AIP Conf. Proc. 1031 (2008) 207-220, [arXiv: 0805.4515 [hep-th]].
\bibitem{P40}
Francesco Bigazzi, Aldo L. Cotrone, Javier Mas, Angel Paredes,
Alfonso V. Ramallo, Javier Tarrio, "D3-D7 Quark-Gluon Plasmas", JHEP
0911(2009)117, [arXiv:0909.2865 [hep-th]].
\bibitem{P41}
A. L. Cotrone, J. M. Pons, P. Talavera, "Notes on a SQCD-like plasma dual and holographic renormalization", JHEP 0711(2007)034, [arXiv:0706.2766 [hep-th]].
\bibitem{P42}
G. Policastro, D. T. Son, A. O. Starinets, "Shear viscosity of strongly coupled N=4 supersymmetric Yang-Mills plasma", Phys. Rev. Lett. 87 (2001) 081601,
[arXiv:hep-th/0104066].
\bibitem{P43}
P. Kovtun, D. T. Son, A. O. Starinets, "Holography and hydrodynamics: diffusion on stretched horizons", JHEP 0310 (2003) 064, [arXiv:hep-th/0309213].
\bibitem{P44}
A. Buchel, J. T. Liu, "Universality of the shear viscosity in supergravity", Phys.Rev.Lett. 93 (2004) 090602, [arXiv:hep-th/0311175].
\bibitem{P45}
P. Kovtun, D. T. Son, A. O. Starinets, "Viscosity in Strongly Interacting Quantum Field Theories from Black Hole Physics", Phys. Rev. Lett. 94 (2005)
111601, [arXiv:hep-th/0405231].
\bibitem{P46}
J. Mas, "Shear viscosity from R-charged AdS black holes", JHEP0603(2006)016, [arXiv:hep-th/0601144]. Kengo Maeda, Makoto Natsuume, Takashi Okamura,
"Viscosity of gauge theory plasma with a chemical potential from AdS/CFT correspondence", Phys. Rev. D73(2006)066013, [arXiv:hep-th/0602010]. J. Sadeghi,
B. Pourhassana, and A. R. Amani, "The effect of higher derivative correction on $\eta/s$ and conductivities in STU model", [arXiv:1011.2291 [hep-th]].
\bibitem{P47}
A. Buchel, "Resolving disagreement for eta/s in a CFT plasma at finite coupling", Nucl. Phys. B803(2008)166-170, [arXiv:0805.2683 [hep-th]].
\bibitem{P48}
A. Buchel, "Shear viscosity of boost invariant plasma at finite coupling", Nucl. Phys. B802(2008)281-306, [arXiv:0801.4421 [hep-th]].
\bibitem{P49}
C. P. Herzog, M. Rangamani and S. F. Ross, "Heating up Galilean
holography", JHEP 0811 (2008) 080,  [arXiv:0807.1099 [hep-th]].
\bibitem{P50}
A. Adams, K. Balasubramanian and J. McGreevy, "Hot Spacetimes for
Cold Atoms", JHEP 0811(2008)059, [arXiv:0807.1111 [hep-th]].
\bibitem{P51}
L. Mazzucato, Y. Oz and S. Theisen, "Non-relativistic Branes", JHEP
0904(2009)073,  [arXiv:0810.3673 [hep-th]].
\bibitem{P52}
E. G. Gimon, A. Hashimoto, V. E. Hubeny, O. Lunin and M. Rangamani,
"Black strings in asymptotically plane wave geometries" JHEP 0308
(2003) 035  [arXiv:hep-th/0306131].
\bibitem{P53}
N. Seiberg and E. Witten, "String theory and noncommutative
geometry" JHEP 9909 (1999) 032,  [arXiv:hep-th/9908142].
\bibitem{P54}
J. M. Maldacena and J. G. Russo, "Large N limit of non-commutative
gauge theo- ries", JHEP 9909 (1999) 025  [arXiv:hep-th/9908134].
\bibitem{P55}
A. Hashimoto and N. Itzhaki, "Non-commutative Yang-Mills and the
AdS/CFT cor- respondence", Phys. Lett. B 465 (1999) 142,
[arXiv:hep-th/9907166].
\bibitem{P56}
Amin Akhavan, Mohsen Alishahiha, Ali Davody, Ali Vahedi, "Non-relativistic CFT and Semi-classical Strings", JHEP 0903(2009)053, [arXiv:0811.3067 [hep-th]].
\bibitem{P57}
C. R. Hagen, "Scale and conformaltransformations in
galilean-covariant field theory", Phys. Rev. D 5 (1972) 377.
\bibitem{P58}
T. Mehen, I. W. Stewart and M. B. Wise, "Conformal invariance for
non-relativistic field theory", Phys. Lett. B 474 (2000) 145 [arXiv:
hep-th/9910025].
\bibitem{P59}
Y. Nishida and D. T. Son, "Nonrelativistic conformal field
theories", Phys. Rev. D 76 (2007) 086004 [arXiv:0706.3746 [hep-th]].
\bibitem{P60}
Sean M. Carroll, Jeffrey A. Harvey, V. Alan Kostelecky, Charles D. Lane, Takemi Okamoto, "Noncommutative Field Theory and Lorentz Violation", Phys. Rev.
Lett. 87 (2001) 141601, [arXiv:hep-th/0105082]. Victor O. Rivelles, "A Review of Noncommutative Field Theories", [arXiv:1101.4579 [hep-th]].
\bibitem{P61}
A. Balachandran, G. Mangano, A. Pinzul, and S. Vaidya,  Int. J. Mod. Phys. A21 (2006) 3111–3126 [arXiv:hep-th/0508002].
\bibitem{P62}
M. Chaichian, M. Sheikh-Jabbari, and A. Tureanu, Phys. Rev. Lett. 86 (2001) 2716 [arXiv:hep-th/0010175]. S. M. Carroll, J. A. Harvey, V. Kostelecky, C. D.
Lane, and T. Okamoto, Phys. Rev. Lett. 87 (2001) 141601 [arXiv:hep-th/0105082].
\bibitem{P63}
J. M. Romero and J. Vergara, Mod. Phys. Lett. A18 (2003) 1673–1680 [arXiv:hep-th/0303064].
\bibitem{P64}
David d'Enterria, "Jet quenching", [arXiv:0902.2011 [nucl-ex]].
\end{thebibliography}
\end{document}